\documentclass[10pt,a4paper,twoside,fleqn,dvips]{article}

\usepackage{amsmath,amssymb}
\usepackage{graphicx}
\usepackage{espcrc2}

\catcode`\á=\active \defá{\'a}
\catcode`\é=\active \defé{\'e} 
\catcode`\É=\active \defÉ{\'E}
\catcode`\è=\active \defè{\`e}
\catcode`\í=\active \defí{\'{\i}}
\catcode`\ó=\active \defó{\'o}
\catcode`\ö=\active \defö{\"o}
\catcode`\Ö=\active \defÖ{\"O}
\catcode`\ú=\active \defú{\'u}
\catcode`\ü=\active \defü{\"u}
\catcode`\Ü=\active \defÜ{\"U}
\catcode`\ñ=\active \defñ{\~n}
\catcode`\Ñ=\active \defÑ{\~N}

\newcommand{\LT}   {\left}
\newcommand{\RT}   {\right}
\newcommand{\eVq}  {\mathrm{eV}^2}
\newcommand{\EtAl} {{\it et al.}}

\title{Analysis of the atmospheric neutrino data in terms of $3\nu$
  oscillations}

\author{M.~Maltoni\address{Instituto de Física Corpuscular~--~C.S.I.C.,
    Universitat de València, \\ Edificio Institutos de Paterna, Apt.~22085,
    E-46071 Valencia, Spain}\thanks{E-mail: \tt maltoni@ific.uv.es}}

\begin{document}
\allowdisplaybreaks

\begin{abstract}
    A global analysis of atmospheric and reactor neutrino data is presented in
    terms of three--neutrino oscillations. We consider in our analysis both
    contained events and upward-going $\nu$-induced muon events, including the
    previous data samples of Fréjus, IMB, Nusex, and Kamioka experiments as
    well as the full 71 kton-yr (1144 days) Super--Kamiokande data set, the
    recent 5.1 kton-yr contained events of Soudan--2 and the results on upgoing
    muons from the MACRO detector. After presenting the results for the
    analysis of atmospheric data alone, we add to our data sample the reactor
    bound of the CHOOZ experiment, showing and important complementarity
    between the atmospheric and reactor limits which results in a stronger
    constraint on the allowed value of $\theta_{13}$.
\end{abstract}

\maketitle

\section{Introduction}

Super--Kamiokande high statistics data~\cite{Gonzalez00,SuperK} indicate a
clear deficit in the $\nu_\mu$ atmospheric events, while $\nu_e$ events appear
to be in good agreement with the Standard Model (SM) prediction. Nowadays, the
present status of two--family oscillations clearly favour the $\nu_\mu \to
\nu_\tau$ channel, while the $\nu_\mu \to \nu_e$ scenario is finally
excluded~\cite{Gonzalez00}; this behaviour is also strengthened by the recent
results of reactor experiments~\cite{CHOOZ}. However, a complete understanding
of the phenomenological picture emerging from the atmospheric neutrino data
require to investigate the whole three--neutrino parameter space.

In this paper we present a global analysis of atmospheric and reactor neutrino
data in terms of three--family neutrino oscillations. We include in our
analysis all the contained events as well as the upward-going neutrino-induced
muon fluxes, including both the previous data samples of Fréjus~\cite{Frejus},
IMB~\cite{IMB}, Nusex~\cite{Nusex} and Kamioka experiments~\cite{Kamiokande}
and the full 71 kton-yr Super--Kamiokande data set~\cite{SuperK}, the recent
5.1 kton-yr contained events of Soudan--2~\cite{Soudan2} and the results on
upgoing muons from the MACRO detector~\cite{MACRO}. We also determine the
constraints implied by the CHOOZ reactor experiments~\cite{CHOOZ}.

In Sec.~\ref{sec:overview} we give a brief introduction to the atmospheric
neutrino problem, describing the data samples, the statistical analysis and
the theoretical calculation of the conversion probabilities for atmospheric
and reactor neutrinos in the framework of three--neutrino mixing. In
Sec.~\ref{sec:analysis} we describe our results for atmospheric neutrino fits,
both by themselves and also in combination with the reactor data. Finally in
Sec.~\ref{sec:sum} we summarize the work and present our conclusions.

\section{Data samples and statistical analysis}
\label{sec:overview}

Underground experiments can record atmospheric neutrinos both by direct
observation of their interactions inside the detector (the so-called {\em
contained events}), and by indirect detection of the muons produced by charged
current interactions in the vicinity of the detector ({\em upgoing--muons
events}).

Contained events have been recorded at six underground experiments, using
either water-Cerenkov detectors (Kamiokande~\cite{Kamiokande}, IMB~\cite{IMB}
and Super--Kamiokande~\cite{SuperK}) or iron calorimeters
(Fréjus~\cite{Frejus}, NUSEX~\cite{Nusex} and Soudan--2~\cite{Soudan2}). For
Kamiokande and Super--Kamiokande, the contained data sample is further divided
into {\em sub-GeV} events, with visible energy below 1.2~GeV, and {\em
multi-GeV} events, with lepton energy above this cutoff. Sub-GeV events arise
from neutrinos of several hundreds of MeV, while multi-GeV events are
originated by neutrinos with energies of several GeV.

Concerning upgoing muons events, in our analysis we considered the latest
results from Super--Kamiokande~\cite{SuperK} and from the MACRO~\cite{MACRO}
experiment. For Super--Kamiokande, the data sample is further divided into
{\em stopping} muons, if the muon stops inside the detector, and {\em
through--going} muons, if the muon track crosses the full detector; for the
MACRO experiment, we only considered through--going events. On average,
stopping muons arise from neutrinos with energies around ten GeV, while
through--going muons are originated by neutrinos with energies around hundred
GeV.

\bigskip

In order to study the results for the different types of atmospheric neutrino
data, we have defined the following combinations of data sets:
\begin{list}{$\bullet$}{
      \setlength{\leftmargin} {4mm}
      \setlength{\parsep}     {0pt}
      \setlength{\itemsep}    {2mm}
      \setlength{\topsep}     {\itemsep}
      \setlength{\partopsep}  {0pt}
      \setlength{\parskip}    {0pt}
      }
  \item FINKS (10 points): the $e$-like and $\mu$-like event rates from the five
    experiments Fréjus, IMB, Nusex, Kamiokande sub-GeV and Soudan--2;
  \item CONT--UNBIN (16 points): the rates in FINKS together with Kamiokande
    multi-GeV and Super--Kamiokande sub- and multi-GeV $e$-like and $\mu$-like
    event rates, without including the angular information;
  \item CONT--BIN (40 points): same as CONT--UNBIN, but also including the
    angular information (5 angular bins) for Kamiokande multi-GeV and
    SK sub- and multi-GeV.
  \item UP--$\mu$ (25 points): muon fluxes for stopping (5 angular bins) and
    through--going (10 bins) muons at Super--Kamiokande and MACRO;
  \item SK (35 points): the full Super--Kamiokande data sample (both contained
    and through--going events);
  \item ALL--ATM (65 points): the full data sample of all atmospheric neutrino
    data.
\end{list}
For each of these combinations we have performed a $\chi^2$ analysis by
computing the $\chi^2$ as a function of the neutrino oscillation parameters.
Following closely the analysis of Refs.~\cite{atm-2,atm-3}, we have used the
actual number of $e$--like and $\mu$--like events, rather than just their
ratios, paying attention both to the correlations between the sources of
errors in the muon and electron predictions and to those amongst the errors of
different energy data samples. Thus we define $\chi^2$ as
\begin{gather} 
    \label{eq:chi2} \chi^2 = \sum_{I,J} \Delta N_I
    \cdot \LT( \sigma_{exp}^2 + \sigma_{th}^2 \RT)_{IJ}^{-1}
    \cdot \Delta N_J, \\
    \sigma_{IJ}^2 \equiv \sigma_\alpha(A)
    \rho_{\alpha\beta}(A,B) \sigma_\beta(B). 
\end{gather}
where $\Delta N_I \equiv N_I^{exp} - N_I^{th}$, $N_I^{th}$ being the predicted
number of events (or the predicted value of the flux, in the case of upgoing
muons) and $N_I^{exp}$ the corresponding experimental measurement. Here $I,J$
stand for any combination of experimental data sets and event-types
considered, {\it i.e.}\ $I = (A,\alpha)$ and $J = (B,\beta)$ where $A,B$ label
the different data samples and $\alpha,\beta$ denote the flavour of the final
lepton. Finally, $\sigma_{IJ}^2$ is the error matrix containing the
theoretical/experimental errors, $\rho_{\alpha\beta}(A,B)$ is the correlation
matrix and $\sigma_\alpha(A)$ is the theoretical/experimental error for the
number of $\alpha$--like events in the $A$ experiment. A detailed discussion
of the errors and correlations used in our analysis can be found in
Ref.~\cite{Gonzalez00}, as well as in the appendices of Refs.~\cite{atm-2},
both for contained and for the upgoing muon data analysis.

\bigskip

In general, the determination of the oscillation probabilities require the
solution of the Schröedinger evolution equation of the neutrino system in
the Earth--matter background. For a CP-conserving three--flavour scenario,
this equation reads
\begin{equation} \label{eq:evol.1}
    i \frac{d\vec{\nu}}{dt} = {\bf H} \, \vec{\nu}, \qquad
    {\bf H} = {\bf R} \cdot {\bf H}_0^d \cdot {\bf R}^\dagger + {\bf V},
\end{equation}
where $\vec{\nu} \equiv \LT( \nu_e, \nu_\mu, \nu_\tau \RT)$, ${\bf R}$ is the
orthogonal matrix connecting the flavour basis and the mass basis in vacuum,
${\bf H}_0^d$ is the vacuum Hamiltonian and {\bf V} describes charged-current
forward interactions in matter:
\begin{gather}
    \label{eq:evol.3}
    {\bf H}_0^d = \frac{1}{2 E_\nu} {\bf diag}
    \LT( - \Delta m^2_{21}, 0, \Delta m^2_{32} \RT), \\
    \label{eq:evol.4} 
    {\bf V} = {\bf diag} \LT( \pm \sqrt{2} G_F N_e, 0, 0 \RT).
\end{gather}
In general, the neutrino transition probabilities depend on five parameters,
namely the two mass-squared differences $\Delta m^2_{21}$, $\Delta m^2_{32}$
and the three mixing angles $\theta_{12}$, $\theta_{13}$, $\theta_{23}$,
however in order to accommodate both solar and atmospheric neutrino data into a
unified framework an hierarchy between $\Delta m^2_{21}$ and $\Delta m^2_{32}$
is required:
\begin{equation} \label{eq:hierarchical}
    \LT.
    \begin{split}
	\Delta m^2_\odot & \equiv \Delta m^2_{21} \\
	\Delta m^2_{atm} & \equiv \Delta m^2_{32}
    \end{split} \RT\} \Rightarrow
    \Delta m^2_{21} \ll \Delta m^2_{32}.
\end{equation}
As a consequence, for what concerns the analysis of atmospheric neutrino
events it is safe to set $\Delta m^2_{21} = 0$, in which case the angle
$\theta_{12}$ also cancels out and the matrix ${\bf R}$ can be written as:
\begin{equation} \label{eq:evolap.1}
    {\bf R} =
    \begin{pmatrix}
	c_{13}          & 0       & s_{13} \\
	- s_{23} s_{13} & c_{23}  & s_{23} c_{13} \\
	- s_{13} c_{23} & -s_{23} & c_{23} c_{13}
    \end{pmatrix}.
\end{equation}
where $c_{ij} \equiv \cos\theta_{ij}$ and $s_{ij} \equiv \sin\theta_{ij}$. As
a result, the $3\nu$ propagation of atmospheric neutrinos can be well
described in terms of only three parameters: $\Delta m^2_{32}$, $\theta_{23}$
and $\theta_{13}$.

When $\theta_{13}=0$, atmospheric neutrinos involve only $\nu_\mu \to
\nu_\tau$ conversions, and in this case there are no matter effects, so that
the solution of Eq.~\eqref{eq:evol.1} is straightforward and the survival
probability takes the well-known vacuum form
\begin{equation} \label{eq:evol.5}
    P_{\mu\mu} = 1 - \sin^2 \LT( 2 \theta_{23} \RT)
    \sin^2 \LT( \frac{\Delta m^2_{32} L}{4 E_\nu} \RT),
\end{equation}
where $L$ is the path-length traveled by neutrinos of energy $E_\nu$. On the
other hand, in the general case of three--neutrino scenario with
$\theta_{13}\neq 0$ the presence of the matter potentials requires a numerical
solution of the evolution equations in order to obtain the oscillation
probabilities for atmospheric neutrinos $P_{\alpha\beta}^\pm$, which are
different for neutrinos and anti-neutrinos because of the reversal of sign in
Eq.~\eqref{eq:evol.4}. In our calculations, we have used for the matter
density profile of the Earth the approximate analytic parameterization of the
PREM model given in Ref.~\cite{Lisi97}.

\bigskip

\begin{figure}[!t]
    \includegraphics[width=0.45\textwidth]{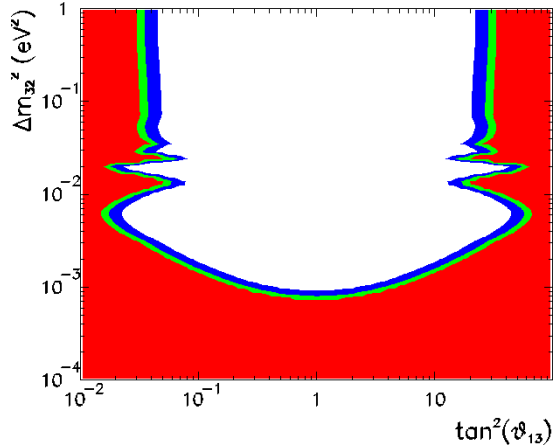}
    \vspace{-8mm}
    \caption{%
      Excluded region (90, 95, 99\% CL) in $\Delta{m}^2_{32}$ and
      $\tan^2\theta_{13}$ from the non observation of oscillations by the
      CHOOZ reactor experiment.} 
    \label{fig:pb.chooz}
    \vspace{-5mm}
\end{figure}

The CHOOZ experiment~\cite{CHOOZ} searches for disappearance of $\bar{\nu}_e$
produced in a power station with two pressurized-water nuclear reactors with a
total thermal power of $8.5$~GW (thermal). At the detector, located at
$L\simeq 1$~Km from the reactors, the $\bar{\nu}_e$ reaction signature is the
delayed coincidence between the prompt ${\rm e^+}$ signal and the signal due
to the neutron capture in the Gd-loaded scintillator. Since no evidence was
found for a deficit of measured vs.\ expected neutrino interactions, the
result of this experiment is a bound on the oscillation parameters describing
the mixing of $\nu_e$ with all the other neutrino species. Under the
assumption $\Delta m_{21}^2 \approx 0$, the only relevant parameters are
$\theta_{13}$ and $\Delta m_{31}^2 \approx \Delta m_{32}^2$, and in
Fig.~\ref{fig:pb.chooz} we show the exclusion plot in this plane. From this
figure we see that at 99\% CL the region $\Delta m_{32}^2 \gtrsim
10^{-3}~\eVq$ and $0.044\lesssim \tan^2\theta_{13}\lesssim 23$ is excluded.

\section{Results of the analysis}
\label{sec:analysis}

\begin{table*}[!t]
    \catcode`?=\active \def?{\hphantom{0}}
    \caption{
      Minimum $\chi^2$ values and best-fit points for various sets of
      atmospheric neutrino data.}
    \label{tab:bestfit}
    \begin{tabular*}{\textwidth}{@{}l@{\extracolsep{\fill}}cccccc}
	\hline
	Data Set & d.o.f.
	& $\tan^2 \theta_{13}$ & $\tan^2 \theta_{23}$
	& $\Delta m_{32}^2~[{\rm eV}^2]$
	& $\chi_{\text{SM}}^2$ & $\chi_{\text{min}}^2$ \\
	\hline
	FINKS       & $10-3$ & $1.???$ & $>100$ & $0.9 \times 10^{-3}$ & $?29.9$ & $14.8$ \\
	CONT--UNBIN & $16-3$ & $0.33?$ & $>100$ & $4.2 \times 10^{-3}$ & $?58.1$ & $18.4$ \\
	CONT--BIN   & $40-3$ & $0.03?$ & $1.6?$ & $2.5 \times 10^{-3}$ & $115.1$ & $41.3$ \\
	SK--CONT    & $20-3$ & $0.03?$ & $0.89$ & $2.3 \times 10^{-3}$ & $?80.7$ & $14.5$ \\
	UP--$\mu$   & $25-3$ & $0.23?$ & $3.3?$ & $7.1 \times 10^{-3}$ & $?50.3$ & $25.1$ \\
	SK          & $35-3$ & $0.005$ & $1.3?$ & $2.8 \times 10^{-3}$ & $131.4$ & $27.8$ \\
	ALL--ATM    & $65-3$ & $0.03?$ & $1.6?$ & $3.3 \times 10^{-3}$ & $191.7$ & $61.7$ \\
	\hline
	ALL--ATM + Chooz & $66-3$ & $0.005$ & $1.4?$ & $3.1 \times 10^{-3}$ &  $191.8$ & $62.5$ \\
	\hline
    \end{tabular*}
\end{table*}

\begin{table*}[!t]
    \catcode`?=\active \def?{\hphantom{0}}
    \caption{Upper bounds on $\theta_{13}$ at 90 and 99\% CL from the analysis of
      the different samples and combinations of atmospheric neutrino data.}
    \label{tab:limits}
    \begin{tabular*}{\textwidth}{@{}l@{\extracolsep{\fill}}ccccc}
	\hline
	Data Set & \multicolumn{3}{c}{$\tan^2\theta_{13}$} 
	& \multicolumn{2}{c}{$\theta_{13}~{\rm [deg]}$} \\
	\cline{2-4} \cline{5-6}
	~           & min     & $90\%$   & $99\%$  & $90\%$       & $99\%$ \\
	\hline
	FINKS       & $1.???$ & $16.4??$ & $42.7?$ & $76.1^\circ$ & $81.3^\circ$ \\
	CONT--UNBIN & $0.33?$ & $10.0??$ & $13.6?$ & $72.5^\circ$ & $74.8^\circ$ \\
	CONT--BIN   & $0.03?$ & $?0.97?$ & $?2.34$ & $44.6^\circ$ & $56.8^\circ$ \\
	UP--$\mu$   & $0.23?$ & $?0.55?$ & $?7.71$ & $36.5^\circ$ & $70.2^\circ$ \\
	SK          & $0.005$ & $?0.33?$ & $?0.68$ & $29.8^\circ$ & $39.5^\circ$ \\
	ALL--ATM    & $0.03?$ & $?0.34?$ & $?0.57$ & $30.1^\circ$ & $37.0^\circ$ \\
	\hline
	ALL--ATM + Chooz & $0.005$ & $?0.043$ & $?0.08$  & $11.7^\circ$ & $15.5^\circ$ \\
	\hline
    \end{tabular*}
\end{table*}

The first result of our analysis refers to the no-oscillation hypothesis. As
can be seen from the fifth column of Table~\ref{tab:bestfit}, the $\chi^2$
values in the absence of new physics clearly show that the experimental data
are totally inconsistent with the SM hypothesis; in fact, the global analysis
(ALL--ATM combination) gives $\chi^2_{SM} = 191.7/\text{(65~d.o.f.)}$,
corresponding to a probability $\lesssim 10^{-14}$. This result is rather
insensitive to the inclusion of the CHOOZ reactor data, as can be seen from
the last line of Table~\ref{tab:bestfit}. Therefore, the Standard Model can be
safely ruled out. In contrast, the $\chi^2$ for the global analysis decreases
to $61.7/\text{(62~d.o.f.)}$ (or $62.5/\text{(63~d.o.f.)}$ including CHOOZ),
acceptable at the 51\% CL, when oscillations are assumed.

\begin{figure}[!t]
    \includegraphics[width=0.45\textwidth]{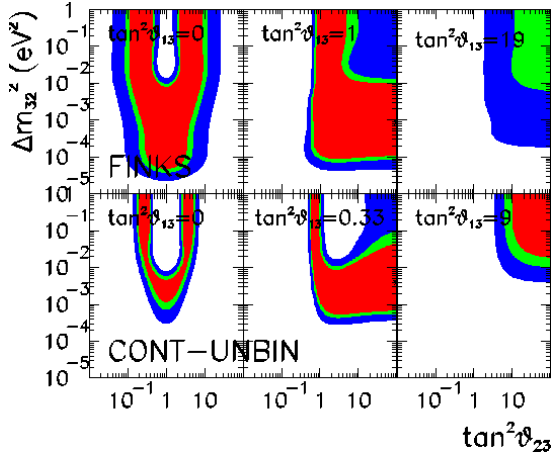}
    \vspace{-8mm}
    \caption{%
      Allowed regions (90, 95, 99\% CL) in $\tan^2\theta_{23}$ and $\Delta
      m^2_{32}$, for different $\tan^2\theta_{13}$ values, for the FINKS and
      CONT--UNBIN combinations.}
    \label{fig:pb.unbin}
    \vspace{-5mm}
\end{figure}

\begin{figure}[!t]
    \includegraphics[width=0.45\textwidth]{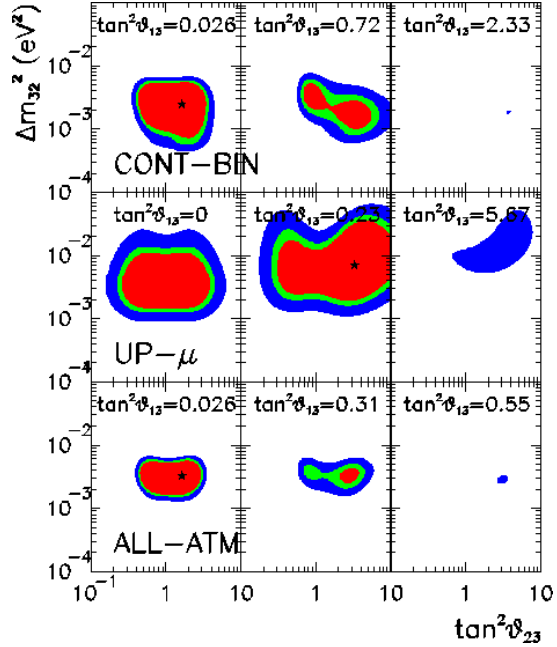}
    \vspace{-8mm}
    \caption{%
      Allowed regions (90, 95, 99\% CL) in $\tan^2\theta_{23}$ and $\Delta
      m^2_{32}$, for different $\tan^2\theta_{13}$ values, for the CONT--BIN,
      UP--$\mu$ and ATM--ALL combinations.}
    \label{fig:pb.bin}
    \vspace{-5mm}
\end{figure}

Table~\ref{tab:bestfit} also gives the minimum $\chi^2$ values and the
resulting best-fit points for the various combinations of data sets
considered. Note that for FINKS, CONT--UNBIN and UP--$\mu$ combinations the
best-fit point is characterized by a rather large value of
$\tan^2\theta_{13}$, while all the other data set favour a value very close to
0. The corresponding allowed regions in the $(\tan^2\theta_{23}, \Delta
m^2_{32})$ plane at 90, 95 and 99\% CL for the different combinations are
depicted in Figs.~\ref{fig:pb.unbin} and \ref{fig:pb.bin}. Note that the value
$\tan^2\theta_{13} = 0$ corresponds to pure $\nu_\mu \to \nu_\tau$
oscillations, and in this case the contour regions are symmetric under the
transformation $\theta_{23} \to \pi/4 - \theta_{23}$ due to the cancellation
of matter effects (cfr.\ Eq.~\ref{eq:evol.5}).

In Fig.~\ref{fig:pb.unbin} we present the allowed regions in
$(\tan^2\theta_{23}, \Delta m^2_{32})$ for different values of
$\tan^2\theta_{13}$, for the FINKS and CONT--UNBIN combinations. It is evident
that, despite of the large statistics provided by Super--Kamiokande, it is not
possible from the information on the total event rates only (i.e.\ without
taking into account the angular dependence) to place any upper bound on
$\Delta m^2_{32}$, and even the lower bound $\Delta m^2_{32} > 2 \times
10^{-4}~\eVq$ is rather weak. Moreover, none of the sets can provide a
relevant constraint neither on $\tan^2\theta_{13}$ nor on $\tan^2\theta_{23}$.

Conversely, in Fig.~\ref{fig:pb.bin} we display the allowed regions in
$(\tan^2\theta_{23}, \Delta m^2_{32})$ for the combinations CONT--BIN,
UP--$\mu$ and ALL--ATM, which also include the information on the angular
distribution of the incoming neutrinos. For what concerns contained events
(CONT--BIN), let us note that the {\it upper} bound on $\Delta m^2_{32}$ is
now rather strong (better than $10^{-2}~\eVq$) as a consequence of the fact
that no suppression for downgoing $\nu_\mu$ neutrinos is observed. This
imposes a lower bound on the neutrino oscillation length and consequently an
upper bound on the mass difference. However contained events alone still allow
values of $\Delta m^2_{32} < 10^{-3}~\eVq$. Note also that the allowed region
is still rather large for $\tan^2\theta_{13} \approx 0.7$ and at 99\% CL it
only disappears for $\tan^2\theta_{13} \simeq 2.4$.

This situation is completely reversed when considering upgoing muon events
(UP--$\mu$). This combination is complementary to CONT--BIN, in the sense that
the corresponding data sets are completely disjoint. In contrast to the
CONT--BIN case, now the upper bound on $\Delta m^2_{32}$ is much weaker
($\approx 3 \times 10^{-2}~\eVq$), while the lower bound is now stronger.
Again, no relevant bound can be put on $\theta_{13}$ from the analysis of
upgoing events alone.

Due to the complementarity between the properties of the CONT--BIN and
UP--$\mu$ combinations, merging them into a single data set leads to much
stronger constraints on the parameter space. In the lower panel of
Fig.~\ref{fig:pb.bin} we show the allowed regions in $(\tan^2\theta_{23},
\Delta m^2_{32})$ from the combined analysis of all the atmospheric neutrino
data (ALL--ATM). Due to the large statistics provided by the Super--Kamiokande
experiment, $\Delta m^2_{32}$ is strongly bounded both from above and from
below. Moreover no region of parameter space is allowed (even at 99\% CL) for
$\tan^2\theta_{13} \gtrsim 0.6$; therefore, this limit can be considered as
the strongest bound which can be put on $\tan^2\theta_{13}$ from the analysis
of atmospheric events alone.

\begin{figure}[!t]
    \includegraphics[width=0.45\textwidth]{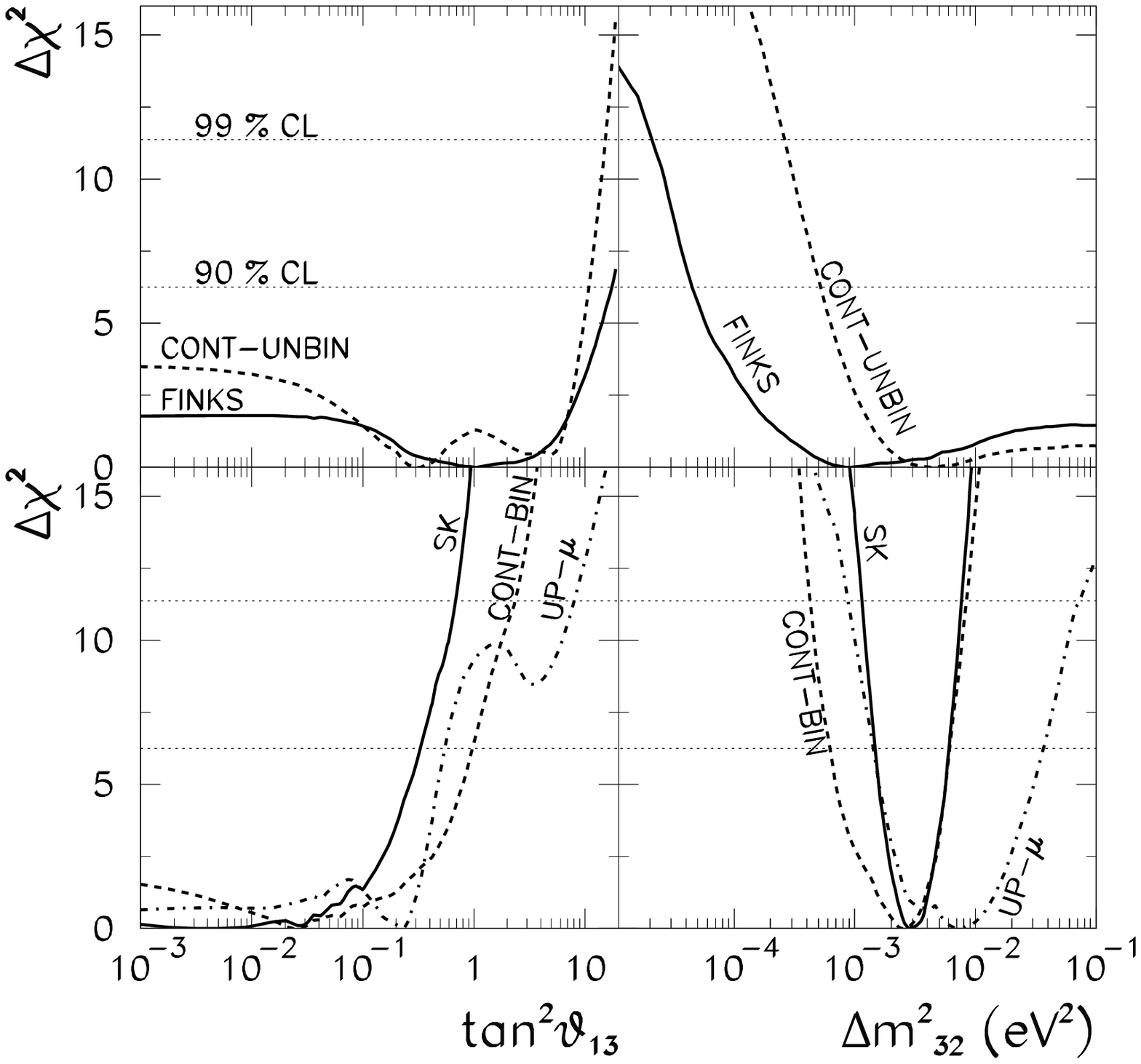} \\[5mm]
    \includegraphics[width=0.45\textwidth]{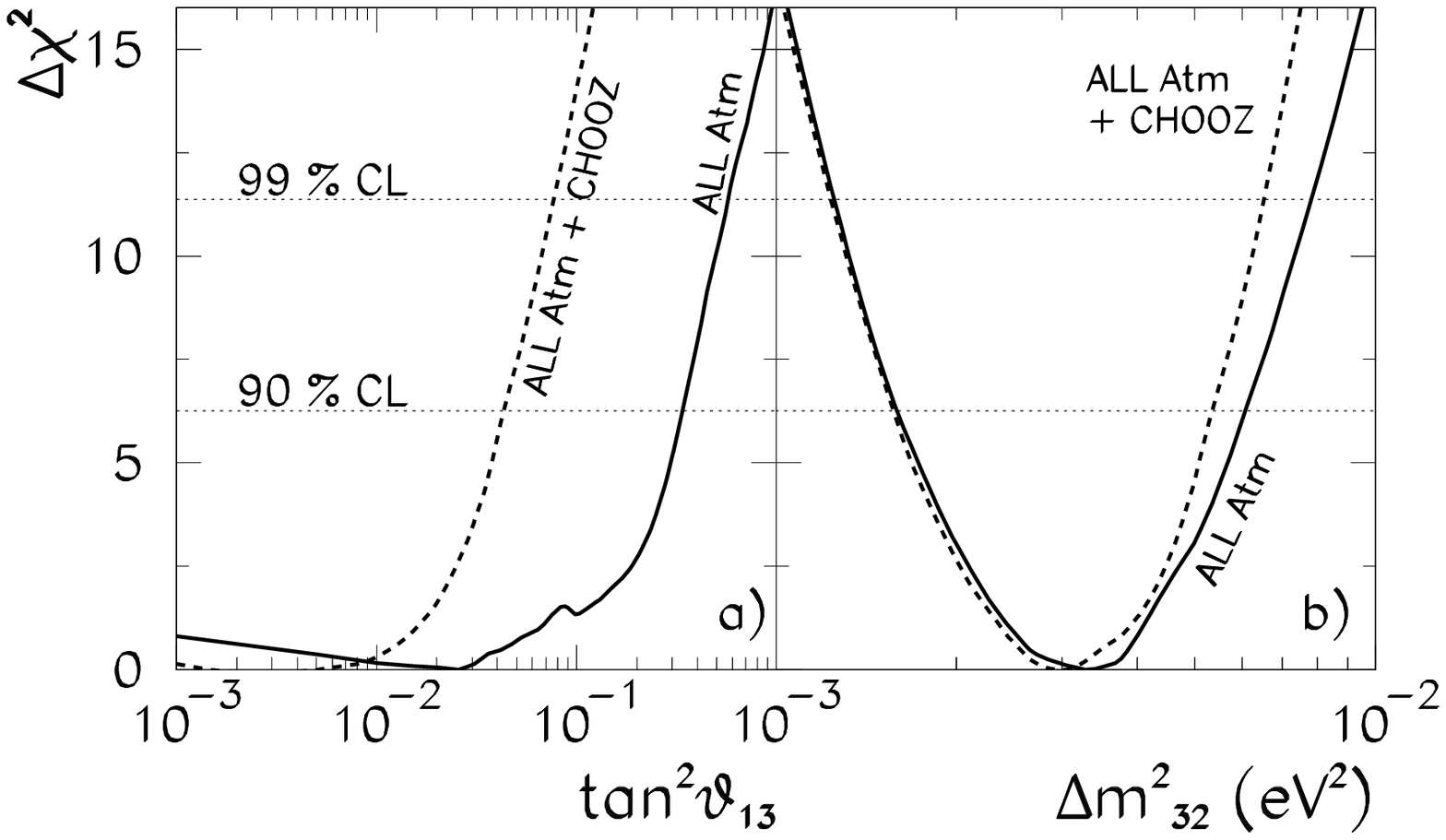}
    \vspace{-8mm}
    \caption{%
      Dependence of $\Delta \chi^2$ on $\tan^2\theta_{13}$ and on $\Delta
      m^2_{32}$, for different combinations of atmospheric neutrino events
      (upper panels) and for the combination ALL--ATM (lower panels, the
      dashed line includes also CHOOZ). The dotted horizontal lines correspond
      to 90 and 99\% CL.}
    \label{fig:chisq}
    \vspace{-5mm}
\end{figure}

All these features can be more quantitatively observed in
Fig.~\ref{fig:chisq}, where we show the dependence of $\Delta\chi^2$ on
$\tan^2\theta_{13}$ and $\Delta m^2_{32}$, for the different combination of
atmospheric neutrino events. In these plots all the neutrino oscillation
parameters which are not displayed have been ``integrated out'', {\it i.e.}\
the displayed $\Delta\chi^2$ is minimized with respect to all the
non-displayed variables. From the left panels of Fig.~\ref{fig:chisq} we can
read the upper bound on $\tan^2\theta_{13}$ that can be extracted from the
analysis of the different samples of atmospheric data alone regardless of the
values of the other parameters of the three--neutrino mixing matrix; the
corresponding 90 and 99\% CL bounds are summarized in Table~\ref{tab:limits}.
Conversely, from the right panels of Fig.~\ref{fig:chisq} we extract the value
of $\Delta m^2_{32}$ allowed by the different combinations, irrespective of
the values of the other mixing parameters.

\begin{figure}[!t]
    \includegraphics[width=0.45\textwidth]{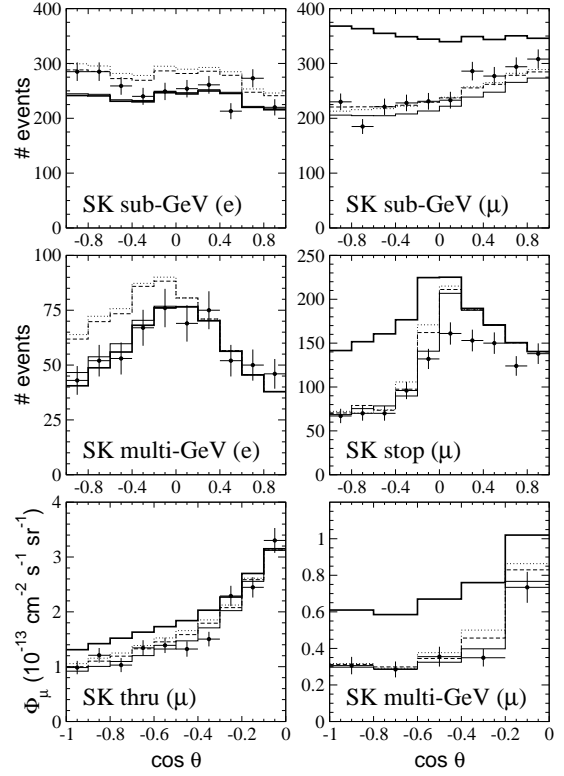}
    \vspace{-8mm}
    \caption{%
      Zenith-angle distributions for Super--Kamiokande events. The thick
      solid line is the expected distribution in the SM, while the thin one
      is the prediction for the overall (ALL--ATM) best-fit point. The
      dashed (dotted) histogram correspond to the distributions for
      $\tan^2\theta_{13} = 0.33\, (0.54)$, $\tan^2\theta_{23} = 3.0\, (3.1)$
      and $\Delta m^2_{32} = 3.3\, (2.85)\times 10^{-3}~\eVq$, which are
      marginally allowed at 90 (99)\% CL.}
    \label{fig:angular}
    \vspace{-6mm}
\end{figure}

In order to illustrate the main effect of varying the angle $\theta_{13}$ in
the description of the angular distribution of atmospheric neutrino events, we
show in Fig.~\ref{fig:angular} the zenith-angle distribution for the
Super--Kamiokande experiment. The thick solid line is the expected
distribution in the absence of oscillation (SM hypothesis), while the thin
full line represents the prediction for the overall best-fit point of the full
atmospheric data set (ALL--ATM, see also Table~\ref{tab:bestfit}) which occurs
at $\tan^2\theta_{13} = 0.03$. The dashed and dotted histograms correspond to
points which are marginally allowed at 90 and 99\% CL, respectively, and occur
at $\tan^2\theta_{13} = 0.33$ and $0.54$ (cfr.\ Table~\ref{tab:limits}). For
each such $\theta_{13}$ value we choose $\Delta m^2_{32} $ and $\theta_{32}$
so as to minimize the $\chi^2$. Clearly the description of the data is
excellent as long as the oscillation is mainly in the $\nu_\mu \to \nu_\tau$
channel (small $\theta_{13}$); this is simply understood since it is clear
from Fig.~\ref{fig:angular} that $e$-like contained events are well accounted
for within the no-oscillation hypothesis. From this figure we see that
increasing $\tan^2\theta_{13}$ leads to an increase in all the contained event
rates; this is due to the fact that an increasing fraction of $\nu_\mu$ now
oscillates as $\nu_\mu\to\nu_e$ (also $\nu_e$'s oscillate as $\nu_e\to\nu_\mu$
but since the $\nu_e$ fluxes are smaller this effect is relatively less
important) spoiling the good description of the $e$--type data, especially for
upgoing multi-GeV electron events. For multi-GeV events all the curves
coincide with the SM one for downgoing neutrinos which did not have the time
to oscillate; this effects is lost in the sub-GeV sample due to the large
opening angle between the neutrino and the detected lepton. We also see that
for multi-GeV electron neutrinos the effect of $\theta_{13}$ is larger close
to the vertical ($\cos\theta=-1$) where the expected ratio of fluxes in the SM
$R(\nu_\mu/\nu_e)$ is larger. Conversely the relative effect of $\theta_{13}$
for $\nu_\mu$ is larger close to the horizontal direction, $\cos\theta=0$.
Concerning upward-going muons events, we see that the effect of adding a large
$\theta_{13}$ in the expected upward muon fluxes is not very significant. For
stopping muons the effect is larger for neutrinos arriving horizontally; this
is due, as for the case of multi-GeV $\mu$-like contained events, to the
larger $R(\nu_e/\nu_\mu)$ SM flux ratio in this direction which implies a
larger relative contribution from $\nu_e$ oscillating to $\nu_\mu$. This
feature is lost in the case of through--going muons because it is partly
compensated by the matter effects and also by the increase of
$\tan^2\theta_{23}$.

In conclusion we see that the analysis of the full atmospheric neutrino data
in the framework of three--neutrino oscillations clearly favours the $\nu_\mu
\to \nu_\tau$ oscillation hypothesis. As a matter of fact the best-fit
corresponds to a small value of $\theta_{13}\approx 9^\circ$, but it still
allows for a non-negligible $\nu_\mu \to \nu_e$ component. More quantitatively
we find that the ranges of parameters are allowed at 99\% CL from this
analysis are (see also Table~\ref{tab:limits}):
\begin{align}
    \Delta m^2_{32}:   && (1.25 & \div 8) \times 10^{-3}~\eVq, \nonumber \\
    \tan^2\theta_{23}: &&  0.37 & \div 6.2, \\
    \tan^2\theta_{13}: &&     0 & \div 0.57. \nonumber
\end{align}
However, one must take into account that these ranges are strongly correlated,
as clearly visible in Fig.~\ref{fig:pb.bin}.

\bigskip

\begin{figure}[!t]
    \includegraphics[width=0.45\textwidth]{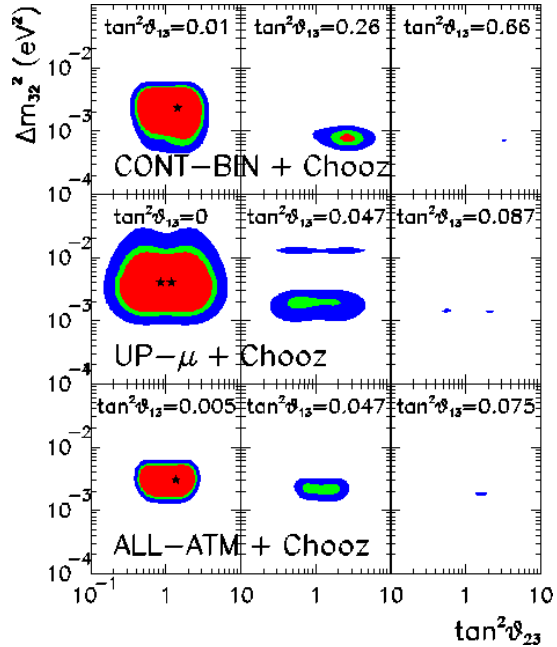}
    \vspace{-8mm}
    \caption{%
      Allowed $(\tan^2\theta_{23}, \tan^2\theta_{13})$ regions for different
      $\Delta m^2_{32}$ values, for the combination ALL--ATM of atmospheric
      neutrino data and including also the CHOOZ result.}
    \label{fig:pb.chbin}
    \vspace{-5mm}
\end{figure}

We now describe the effect of including the CHOOZ reactor data together with
the atmospheric data samples in a combined three--neutrino $\chi^2$ analysis,
under the hypothesis $\Delta m^2_{21}\lesssim 3\times 10^{-4}~\eVq$. The
results of this analysis are summarized in Figs.~\ref{fig:chisq}
and~\ref{fig:pb.chbin}, as well as in the last line of Tabs.~\ref{tab:bestfit}
and~\ref{tab:limits}. As already discussed, the negative results of the CHOOZ
reactor experiment strongly disfavour the region $0.044\lesssim
\tan^2\theta_{13}\lesssim 23$ if $\Delta m^2_{32}\gtrsim 10^{-3}~\eVq$,
however for smaller values of $\Delta m^2_{32}$ the corresponding bound on
$\theta_{13}$ is much weaker. Therefore an independent lower bound on $\Delta
m^2_{32}$ is required for the CHOOZ constraint on $\theta_{13}$ to be
applicable. To illustrate this point we show in the upper panels of
Fig.~\ref{fig:pb.chbin} the allowed regions from the combination of the
CONT--BIN events with the CHOOZ data. Comparing them with the corresponding
ones from Fig.~\ref{fig:pb.bin}, we see that, although the region $\Delta
m^2_{32} > 10^{-3}~\eVq$ is ruled out as soon as $\theta_{13}$ deviates from
zero, there is still a part of the parameter space which survives (at 99\% CL)
up to $\tan^2\theta_{13} \approx 0.66$. Thus, even with the large statistics
provided by the Super--Kamiokande data and including the CHOOZ result, it is
not possible to improve the bound on $\theta_{13}$ using only contained
events.

The situation is changed once the upgoing muon events are included in the
analysis. As clearly visible from the central panels of Fig.~\ref{fig:pb.bin},
the UP--$\mu$ data sample disfavours the low mass region $\Delta m^2_{32} <
10^{-3}~\eVq$. As a result, the full 99\% CL allowed parameter regions from
the global analysis of the atmospheric data ALL--ATM lies in the mass range
where the CHOOZ experiment should have observed oscillations for sizeable
$\theta_{13}$ values, and this leads to the shift of the global minimum from
the combined atmospheric plus CHOOZ data to $\theta_{13} \approx 0$. Thus
adding the reactor data has as main effect the strong improvement of the
$\tan^2\theta_{13}$ limit, as can also be seen from the lower-left panel in
Fig.~\ref{fig:chisq}. Conversely, from the lower-right panel of the same
figure we see that the allowed range of $\Delta m^2_{32}$ is almost
unaffected.

In conclusion, from the atmospheric and reactor neutrino combined analysis we
get the following 99\% CL allowed ranges of parameters:
\begin{align}
    \Delta m^2_{32}:   && (1.2 & \div 6.6) \times 10^{-3}~\eVq, \nonumber \\
    \tan^2\theta_{23}: && 0.36 & \div 3.3, \\
    \tan^2\theta_{13}: &&    0 & \div 0.08. \nonumber
\end{align}

\section{Summary and conclusions}
\label{sec:sum}

In this article we have presented a global analysis of both atmospheric and
reactor neutrino data in terms of three--flavour oscillations. Our results
show that the most favourable scenario is the $\nu_\mu \to \nu_\tau$
oscillation hypothesis, however from the analysis atmospheric data alone it is
not possible to derive a strong bound on $\theta_{13}$, and consequently a
non-negligible $\nu_\mu \to \nu_e$ component is still allowed. When the CHOOZ
data in included into the analysis, the bound on $\theta_{13}$ is drastically
improved, and $e$--like neutrinos completely decouple from the oscillation
picture. Conversely, the preferred mixing in the $\nu_\mu \to \nu_\tau$
channel -- described by the angle $\theta_{23}$ -- appear to be nearly
maximal. Together with the result of solar experiments, which also prefer
maximal mixing in $\theta_{12}$ (although in this case the result is still not
overwhelming), there seems to be an indication that the ultimate structure of
neutrino mixing is bi-maximal. This result is in puzzling contrast with the
observed structure of quark mixing, and may indicate that the origin of
neutrino masses is intrinsically different from the standard Higgs mechanism
which is responsible for the masses of the charged leptons in the Standard
Model.

\section*{Acknowledgments}

I wish to thank my collaborators M.C.\ Gonzalez-Garcia, C.\ Peña-Garay and
J.W.F.\ Valle. This work is supported by DGICYT under grant PB98-0693, by the
European Commission TMR networks ERBFMRX-CT96-0090 and HPRN-CT-2000-00148, and
by the European Science Foundation network N.~86.

\end{document}